\documentclass{ijcaArticle}

\usepackage[none]{hyphenat}

\ijcaVolume{176}
\ijcaNumber{24}
\ijcaYear{2020}
\ijcaMonth{May}

\begin{document}

\title{Eye Movements Biometrics: a Bibliometric Analysis from 2004 to 2019} 

\author{ 
   \large Antonio Ricardo Alexandre Brasil \\[-3pt]
   \normalsize Instituto Federal do Espírito Santo  \\[-3pt]
    \normalsize ES-010 Km-6,5, 29173-087 \\[-3pt] 
    \normalsize Serra, Espírito Santo, Brazil \\[-3pt] 
    \normalsize	anribrasil@gmail.com \\[-3pt]
  \and
   \large Jefferson Oliveira Andrade  \\[-3pt]
   \normalsize Instituto Federal do Espírito Santo  \\[-3pt]
    \normalsize ES-010 Km-6,5, 29173-087 \\[-3pt] 
    \normalsize Serra, Espírito Santo, Brazil \\[-3pt] 
    \normalsize	jefferson.andrade@ifes.edu.br \\[-3pt]
\and
   \large Karin Satie Komati  \\[-3pt]
   \normalsize Instituto Federal do Espírito Santo  \\[-3pt]
    \normalsize ES-010 Km-6,5, 29173-087 \\[-3pt] 
    \normalsize Serra, Espírito Santo, Brazil \\[-3pt] 
    \normalsize	kkomati@ifes.edu.br  \\[-3pt]
}

\keywords{eye movements, person identification, bibliometric analysis, biometrics}

\maketitle

\begin{abstract} 
  Person identification based on eye movements is getting more and more attention, as it is anti-spoofing resistant and can be useful for continuous authentication. Therefore, it is noteworthy for researchers to know who and what is relevant in the field, including authors, journals, conferences, and institutions. This paper presents a comprehensive quantitative overview of the field of eye movement biometrics using a bibliometric approach. All data and analyses are based on documents written in English published between 2004 and 2019. Scopus was used to perform information retrieval. This research focused on temporal evolution, leading authors, most cited papers, leading journals, competitions and collaboration networks.
\end{abstract}

\section{Introduction}

The idea of using eye movements to identify people was first reported by the pioneering work of Kasprowski and Ober \cite{kasprowski2004eye}. In their experiments, jumping dots displayed on the screen in specific intervals were used as a stimulus for the users. For a database of nine persons, the best result was an average false acceptance rate of 1.48\% and an average false rejection rate of 22.59\%.  The popularity of eye movement as biometrics has grown for many reasons: it is resistant to anti-spoofing \cite{komogortsev2015attack, rigas2015eye} and can be used for continuous authentication of the user \cite{niinuma2010soft, crouse2015continuous, kinnunen2010towards}, as well as capturing non-invasive movements.  

Currently, many institutions and researchers have been studying the potential of eye movements biometrics for the future and a series of competitions have been held to stimulate research in the field. To identify the evolution in the field, a quantitative approach, the bibliometric analysis, was chosen for the present study. 

The bibliometric methodology was developed based on the statistical bibliography proposed by Hulme et al.  \cite{hulme1923statistical},  later developed by Pritchard et al. \cite{pritchard1969statistical}, and relies on the quantitative study of bibliographic records. It is part of librarianship and information science being used in several aspects like the selection of books and periodicals, the evaluation of bibliographies, and historical applications \cite{lawani1981bibliometrics}.  

This article will provide useful information for researchers in the field of eye movement biometrics. To achieve this goal, this article will focus on the analysis of all records since the first paper about eye movement biometrics. The analysis will be on the 189 articles written in English and published from 2004 to 2019, indexed by Scopus.

This paper is organized as follows: Section 2 describes the methodological approach of this research. Section 3 presents the delimitations of the study. The results obtained and discussions of the present study in Section 4. Competitions in eye movement biometrics in Section 5. Finally, Section 6 presents the conclusion of our bibliometric study.

\section{Methodological approach}

To identify the development in the field of eye movement biometrics, this paper used the  Scopus Database (created by Elsevier) \cite{elsevier2019a}. A  retrospective search on Scopus in the period 2004-2019. The search was performed using the  query: \textbf{TITLE-ABS-KEY (``biometric*''  OR  ``person identification''  OR  ``person recognition'')  AND  TITLE-ABS-KEY (``eye movement''  OR  ``sacca*''  OR  ``ocular movement''  OR  ``gaze track*'')  AND  PUBYEAR  $>$  2003 AND PUBYEAR $<$ 2020 }. 

By using these terms,  242 documents were found, being 240 in English, 1 in Chinese, and 1 in Turkish. Only documents written in English were used, considering the overall scope of the language. The distribution of the 240 documents per type is:  Conference Paper (125; 52.08\%), Article (86; 35,83\%), Conference Review (19; 7.92\%), Book Chapter (6; 2.50\%),  Review (4; 1.67\%). Documents with the type of Conference Review and Book Chapter were removed from the bibliometric analysis, just as Alvarez-Betancourt and Garcia-Silvestre \cite{alvarez2014overview} did, reaching 215 documents for analysis. 

\begin{figure*}[htb]
\begin{center}
\includegraphics[width=0.7\linewidth]{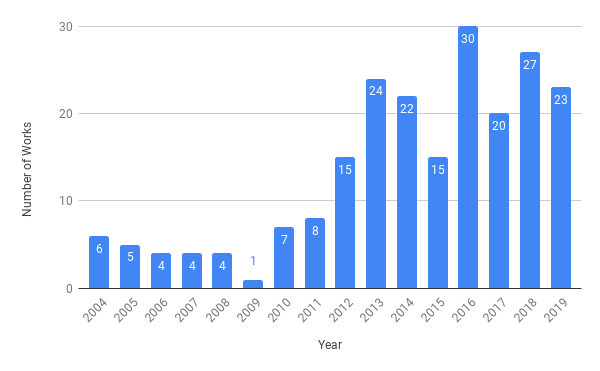}
\end{center}
   \caption{Articles using eye movement biometrics for each year, from 2004 to 2019. The x-axis shows the years from 2004 to 2019 and y-axis shows the number of publications.}
\label{fig:works1}
\end{figure*}

Since the Scopus database is updated daily, all search results have been exported to the CSV file to save the results for future reference. Thereafter, a process of normalizing the names of the authors was performed, due to cases of duplicate author names (e.g. the author Oleg Komogortsev is sometimes just O. Komogortsev). In case the previous procedure failed to resolve ambiguities, additional information was retrieved from the profile of Google Scholar to clear any doubts.



It is not a qualitative study describing methods and techniques, neither a state-of-art of eye movement biometrics research. In this sense, an adequate methodological approach is necessary to identify in the current literature the promising of the research. For a study of the state-of-art, this work highly recommends reading the article \cite{rigas2017current}. The result of the search in the Scopus database brought articles that are not about eye movement like biometrics, although they are related themes. Some articles have even been listed as most cited in Table \ref{tab:tabela_5}, but are marked with '*'.



\section{Results and discussion}

The temporal evolution of the number of works in the field is  presented in Figure \ref{fig:works1}. It is possible to see a solid and incremental amount of publications.  The x-axis shows the years from 2004 to 2019 and the y-axis shows the numbers of publications. In the period of 2006 to 2008, there was no increase in the number of papers, but in the period of 2010 to 2013, it is possible to observe a solid growth in the number of articles. The blue line shows the growth trend in the years from 2004 to 2019.

\subsection{Leading authors}

From the 215 papers selected, 553 authors were identified, of which 22.60\% published more than one paper. Statistical data from the 20 authors with the highest amount of production are shown in  Table \ref{tab:tabela_x}. The first column is the ranking classified by the number of publications of the author; the second column is the name of the author; in the third column, the number of publications (P); the fourth column  shows the number of collaborations (C) and; in the fifth column, the number of times the author was cited (R). 

\begin{table*}[t!]
\tbl{The 20 most productive authors, where P is the number of publications, C the number of collaborations and R the number of times that the author has been cited. Rank is given by R divided by P. The Betwe., Clos. and Deg. refers respectively to Betweenness, Closeness and, Degree. The last column presents all publications references' for each author. The top three values in each column are in bold.}{
\centering

\begin{tabular}{lcccccccccp{5.8cm}}
\hline
\multicolumn{1}{c}{Author} & P  & C  & R  & R/P    & Rank & h-index   & Betwe.       &  Clos.     & Deg.   		& References \\ \hline
1. Komogortsev, O. V.      & \textbf{34} & \textbf{21} & \textbf{94} & 2.76   & 18   & \textbf{14}        & \textbf{0.03}     & \textbf{0.14}   & \textbf{0.29} 	&  \cite{friedman2019assessment,lohr2018implementation,zemblys2018using,rigas2018study,friedman2017method,rigas2017current,rigas2016towards,abdulin2016eye,komogortsev2016oculomotor,rigas2016biometric,abdulin2015person,komogortsev2015bioeye,rigas2015eye,komogortsev2015attack,rigas2014biometricpspem,rigas2014biometric,komogortsev2014application,komogortsev2014template,rigas2014gaze,komogortsev2015biometrics,holland2014software,holland2013complex,komogortsev20132d,komogortsev2013liveness,holland2013complexb,brooks2013perceptions,komogortsev2013biometric,kasprowski2012first,holland2012biometric,komogortsev2012multimodal,komogortsev2012biometric,komogortsev2012cue,holland2011biometric,komogortsev2010biometric} \\
2. Holland, C. D.          & \textbf{14} & \textbf{9}  & \textbf{75} & 5.35   & 7    & \textbf{8}         & 0.00     & 0.08   & 0.03	& \cite{komogortsev2016oculomotor,komogortsev2015attack,komogortsev2014application,komogortsev2014template,komogortsev2015biometrics,holland2014software,holland2013complex,komogortsev20132d,holland2013complexb,komogortsev2013biometric,holland2012biometric,komogortsev2012multimodal,komogortsev2012cue,holland2011biometric}  \\
3. Rigas, I.               & \textbf{14} & 4  & \textbf{58} & 4.14   & 10   & \textbf{8}         & \textbf{0.01}     & \textbf{0.09}   & \textbf{0.16}	& \cite{rigas2018study,rigas2017current,rigas2016towards,abdulin2016eye,rigas2016biometric,komogortsev2015bioeye,rigas2015eye,rigas2014biometricpspem,rigas2014biometric,rigas2014gaze,kasprowski212influence,kasprowski2013influence,rigas2012human,rigas2012biometric}  \\
4. Kasprowski, P.          & 13 & 2  & 58 & 4.46   & 9    & 8         & \textbf{0.00}     & 0.01   & \textbf{0.10}	& \cite{kasprowski2018fusion,kasprowski2016disk,kasprowski2016using,harezlak2017eye,kasprowski2014second,kasprowski212influence,kasprowski2013impact,kasprowski2013influence,kasprowski2012first,kapczynski2006modern,kasprowski2005enhancing,kasprowski2004eye,kasprowski2004flick} \\
5. Karpov, A.              & 10 & \textbf{10} & 60 & 6      & 4    & 8         & 0            & \textbf{0.10}   & 0.03	&  \cite{komogortsev2016oculomotor,komogortsev2015attack,komogortsev2014template,komogortsev2015biometrics,komogortsev20132d,komogortsev2013liveness,kasprowski2012first,komogortsev2012multimodal,komogortsev2012biometric,komogortsev2012cue}	\\
6. Juhola, M.              & 6  & 5  & 27 & 4.5    & 8    & 4         & 0            & 0.01   & 0.05	& \cite{zhang2017biometrics,zhang2015biometric,zhang2014biometric,zhang2014biometricc,zhang2013applying,juhola2013biometric,zhang2012biometric} \\
7. Zhang, Y.               & 6  & 1  & 57 & \textbf{9.5}    & 3    & 3         & 0.00     & 0.01   & 0.07	& 	\cite{zhang2017biometrics,zhang2015biometric,zhang2014biometric,zhang2014biometricc,zhang2013applying,juhola2013biometric,zhang2012biometric}	 \\
8. Martinovic, I.          & 6  & 6  & 34 & 5.6    & 5    & 4         & 0            & 0.03   & 0.03	& 	\cite{eberz201928,eberz2018your,sluganovic2018analysis,eberz2017evaluating,sluganovic2016using,eberz2016looks}		 \\
9. Deravi, F.              & 5  & 4  & 27 & 5.4    & 6    & 3         & 0.00     & 0.01   & 0.03	& 	\cite{ali2019gaze,ali2018gaze,ali2013spoofing,ali213spoofing,deravi2011gaze}		 \\
10. Harezlak, K.           & 5  & 4  & 15 & 3      & 15   & 3         & 0.00     & 0.01   & 0.05 	&  \cite{kasprowski2018fusion,kasprowski2016disk,kasprowski2016using,harezlak2017eye,kasprowski2014second}     \\
11. Rasmussen, K.B.        & 5  & 5  & 19 & 3.8    & 12   & 3         & 0            & 0.03   & 0.03	&  	\cite{eberz201928,sluganovic2018analysis,eberz2017evaluating,sluganovic2016using,eberz2016looks} \\
12. Ali, A.                & 4  & 4  & 13 & 3.25   & 13    & 2         & 0            & 0          & 0.05	& 	\cite{ali2019gaze,ali2018gaze,ali2013spoofing,ali213spoofing}		 \\
13. Eberz, S.              & 4  & 0  & 16 & 4      & 11   & 3         & 0            & 0          & \textbf{0.10}	&  	\cite{eberz201928,eberz2018your,eberz2017evaluating,eberz2016looks}	 \\
14. Hoque, S.              & 4  & 4  & 12 & 3      & 15   & 2         & 0            & 0.01   & 0.01	&  \cite{ali2019gaze,ali2018gaze,ali2013spoofing,ali213spoofing}			 \\
15. Lenders, V.            & 4  & 4  & 13 & 3.25   & 13    & 3         & 0            & 0.01   & 0.01	& \cite{eberz201928,eberz2018your,eberz2017evaluating,eberz2016looks}			 \\
16. Friedman, L.           & 3  & 3  & 9  & 3      & 15    & 2         & 0.00     & 0.06  & 0.05	&  	\cite{friedman2019assessment,rigas2018study,friedman2017method}		 \\
17. Hansen, D. W.          & 3  & 2  & 41 & \textbf{13.7}  & 2   & 3         & 0.00     & 0.03   & 0.05	& \cite{batista2015depth,vitonis2014person,hansen2010eye} 			 \\
18. Ober, J.               & 3  & 3  & 89 & \textbf{29.7}  & 1    & 3         & 0            & 0.02   & 0.01	&  	\cite{kasprowski2005enhancing,kasprowski2004eye,kasprowski2004flick}		 \\
19. Saeed, U.              & 3  & 0  & 4  & 1.33   & 19   & 2         & 0            & 0          & 0			&  	\cite{saeed2016eye,saeed2014survey,saeed2014automatic}		 \\
20. Wang, X.               & 3  & 2  & 2  & 0.66   & 20    & 2         & 0.00     & 0.03   & 0.07	&  	\cite{ma2018integrating,wang2017face,wang2015survey,wang2007novel}	 \\
\end{tabular}}
\label{tab:tabela_x}
\end{table*}

In terms of the number of publications (P), the author with most publications is the researcher Oleg Komogortsev, professor at Texas State University, with 34 papers about the research field, twice as much as the numbers of papers of the second author. There are four more authors with more than ten papers: Ioannis Rigas (researcher at Gemalto Cogent Inc.); Pawel Kasprowski (professor at Politechnika Slaska); Corey D. Holland and Alex Karpov (Texas State University). The publication of these five authors sums 85 works, representing 39.53\% of 215 works.
Evaluating the number of collaborations (column C), researcher Oleg Komogortsev maintains first place with 21 collaborations, followed by researcher Karpov with 10 collaborations and Corey D. Holland with 9 collaborations. A detailed network analysis will be presented in the next section.

The most cited authors are (column R): Oleg Komogortsev (with 94 citations in his 34 works), Pawel Kasprowski (with 58 citations) and Corey D. Holland (with 58 citations). The sixth column of Table \ref{tab:tabela_x} shows the division (R/P) and; the seventh is the Rank based on the R/P rate. In this last column,  position 2 repeats in the last two rows, which means that both authors are in the second position, according to R/P. When the repetition occurs, the next position is increased by one, in this case, skipping position number 3. In this simplified ranking, taking into the total number of publications divided by the total number of citations, the three authors are Jósef Ober (professor at the Silesian University of Technology); Dan Witzner Hansen (professor at the University of Copenhagen) and Youming Zhang.

Nevertheless, the index (R/P) is not so effective. For example, the authors Jósef Ober and Dan Witzner Hansen have three papers with a high number of citations. For Jósef Ober, these three papers were written in collaboration with Pawel Kasprowski, who had more publications and Dan Witzner Hansen published a survey of models for eyes and gaze, with high number of citations. Thus, for the relevance of publications, the h-index will be used. It is defined as the highest number of publications of a scientist that received 'h' or more citations each while the other publications have not more than 'h' citations each  \cite{hirsch2005index}. In this work, only the articles selected by the Scopus search will be considered. Thus, the h-index calculated by Google Scholar or another website will not be used, because in the profile of these websites the author's h-index is calculated considering all the  works published, regardless if it is about eye movements biometrics or not.


The calculated h-index for each author is shown in the eighth column of Table \ref{tab:tabela_x}. The author with the highest h-index is Oleg Komogortsev with an h-index of 14, followed by Holland, Rigas, Kasprowski, and Karpov with 8. 

\subsection{Collaboration network}

\begin{figure*}[h!]
\begin{center}
\includegraphics[width=\linewidth]{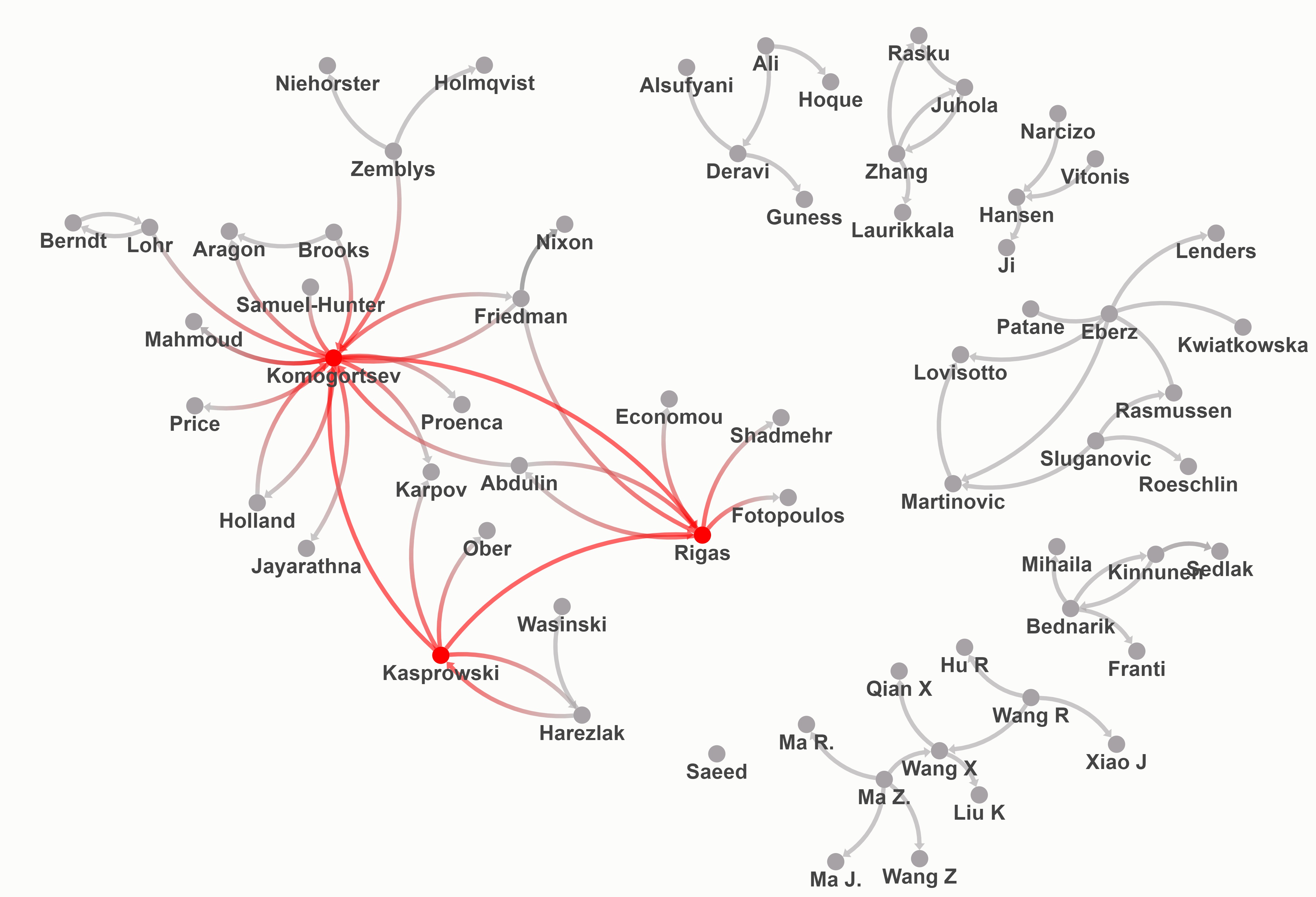}
\end{center}
   \caption{Collaboration network of works related to eye movement biometrics. Each node represents an author and each vertex represents the joint publication between the authors (nodes).}
\label{fig:figura1}
\end{figure*}

\begin{table*}[h!]
\tbl{Top 15 most cited papers retrieved from Scopus.}{
\begin{tabular}{|c|p{13cm}|c|c|}
\hline
Rank & Title & Year & Cited by \\
\hline\hline
      1 & In the Eye of the Beholder: A Survey of Models for Eyes and Gaze* \cite{hansen2010eye} & 2010 & 835  \\
      2 & Eye movements in biometrics \cite{kasprowski2004eye} & 2004 & 86  \\
      3 & Clinical criteria for subtyping Parkinson's disease: Biomarkers and longitudinal progression* \cite{fereshtehnejad2017clinical} & 2005 & 81  \\
      4 & Eye-movements as a biometric \cite{bednarik2005eye} & 2005 & 79  \\
      5 & Ocular biometrics: A survey of modalities and fusion approaches \cite{journals/inffus/NigamVS15} & 2015 & 73 \\
      6 & Biometric identification via eye movement scanpaths in reading \cite{holland2011biometric} & 2011 & 72  \\
      7 & The use of rodent skilled reaching as a translational model for investigating brain damage and disease* \cite{klein2012use} & 2012 & 63  \\
      8 & Biometric identification based on the eye movements and graph matching techniques \cite{rigas2012biometric} & 2012 & 59  \\
      9 & Multitask learning for EEG-based biometrics* \cite{sun2008multitask} & 2008 & 54  \\
      10 & Towards task-independent person authentication using eye movement signals \cite{kinnunen2010towards} & 2010 & 53  \\
      11 & Unaware Person Recognition From the Body When Face Identification Fails* \cite{rice2013unaware} & 2013 & 52 \\
      12 & Biometric authentication via oculomotor plant characteristics \cite{komogortsev2012biometric} & 2012 & 50  \\
      13 & Biometric identification via an oculomotor plant mathematical model \cite{komogortsev2015biometrics} & 2010 & 47 \\
      14 & Complex eye movement pattern biometrics: Analyzing fixations and saccades \cite{holland2013complex} & 2013 & 43 \\
      15 & Cross-database face antispoofing with robust feature representation* \cite{patel2016cross} & 2016 & 42 \\
\hline
\end{tabular}}
\label{tab:tabela_5}
\end{table*}

\begin{table*}[h!]
\tbl{Top 3 journals with most published works.}{
\begin{tabular}{|p{13cm}|c|c|}
\hline
Name & IF & \#  \\
\hline\hline
1. Proceedings of SPIE - The International Society for Optical Engineering \cite{spie2019} & 0.993 & 11 \\
2. Pattern Recognition Letters \cite{prl2019} &  1.952 & 6\\
3. IEEE Transactions on Information Forensics and Security \cite{ieeetifs2019} & 5.824 & 6 \\
\hline
\end{tabular}}
\label{tab:tabela_3}
\end{table*}

\begin{table*}[h!]
\tbl{Top 5 conventions with the most publications.}{
\begin{tabular}{|p{16cm}|c|}
\hline
Name & \#  \\
\hline\hline
1. Lecture Notes in Computer Science (including subseries Lecture Notes in  Artificial Intelligence and Lecture Notes in Bioinformatics) \cite{lncs2019} &  15 \\
2. Eye Tracking Research and Applications Symposium (ETRA) \cite{etra2019} & 5  \\
3. International Journal of Biometrics \cite{ijb2019} &  4 \\
4. 2012 IEEE 5th International Conference on Biometrics: Theory, Applications and Systems, BTAS 2012 \cite{ieeebiometrics2019} &  4  \\
5. Advances in Intelligent Systems and Computing \cite{acmins2019}     & 3  \\
\hline
\end{tabular}}
\label{tab:tabela_4}
\end{table*}

To evaluate the collaborations between the authors, this work used an analysis based on a network of collaborations. Each node represents an author and each vertex represents the joint publication between the authors (nodes). The network of collaborations is presented in Figure \ref{fig:figura1}. To generate the network, the authors of Table \ref{tab:tabela_x} and all their collaborators were selected -- regardless of whether or not the author is in the list of most productive authors.


For the collaboration analysis, this work obtained the centrality measures to identify the pattern of the most productive authors \cite{yan2010mapping, alvarez2014overview}. These metrics are: betweenness (column Betwe. of Table \ref{tab:tabela_x}), closeness (column Clos. of Table \ref{tab:tabela_x}) and degree (column Deg. of Table \ref{tab:tabela_x}). 

 Degree centrality aims to measure the number of direct connections of a node within the network, being important to identify the collaboration network. The degree is the easiest measure to see in Figure \ref{fig:figura1} because a degree is the direct connection of the node and indicates the number of collaborations (C) shown in Table \ref{tab:tabela_x}. The nodes with the highest degree are in red in Figure \ref{fig:figura1},  Kasprowski, Komogortsev and Rigas, which cooperate with each other in the network. The degree is normalized by dividing by the maximum possible degree in the graph $(N-1)$ where N is the number of nodes in G. 

 Betweenness is a measure that, given all the nodes (authors), calculates the number of minimum paths using a geodesic distance from each node to all other nodes that connect through it. This metric aims to analyze the influence of an author to participate in the network, being a bridge between the parts of the network. Closeness measures the distance from one vertex to all other vertices. In this case, it is the distance from an author along with all other authors using a geodesic distance. It is an important metric to identify if an author is efficient, ie, if the author collaborates with other authors in the network. Closeness and betweenness values were normalized, considering the number of nodes in the connected part of the graph containing the node.
 
A different row of Table \ref{tab:tabela_x} refers to the author Saeed, whose metrics have betweenness, closeness, and degree with zero values. One of the reasons that this author has zero values in his metrics is because this author does not have any collaboration in the network, a fact that is possible to see in Figure \ref{fig:figura1}.  

Another analysis that can be made is the author Holland has a  high h-index, but zero in betweenness. This is due to the author has many collaborations with Oleg Komogortsev, but no collaborations with other authors of the network. Ober, for example, has a closeness value of 0.0242, but zero betweenness, this is due Ober have been working only with Kasprowski and he has no interactions with other authors. The author with the highest metrics of collaborations is Oleg Komogortsev.


In the analysis of the collaboration network between institutions, there is the strong presence of Texas State University, with a large number of works, which is the institution of author Oleg Komogortsev.



\subsection{Most cited papers}

We made an analysis of the most cited articles retrieved from Scopus. Some papers retrieved in the Scopus search were not about eye movement biometrics, but are present in Table \ref{tab:tabela_5} with '*'. 

The first study of eye movement as biometric identification \cite{kasprowski2004eye} is present in the table as Rank 2. It is possible to correlate the most cited papers with the most productive authors, for example, the fourth most productive authors Kasprowski, Komogortsev, Rigas and Holland, in Table \ref{tab:tabela_x} present the works most cited papers in the ranking, presented respectively in the ranking 2, 6, 8 and 12.


\subsection{Leading  journals and conventions}

Considering all the collected works, a study was conducted to evaluate the journals and conventions that concentrated most publications. Table \ref{tab:tabela_3} lists the top 3 journals and Table \ref{tab:tabela_4} top 5 conventions, ranked in descending order. The impact factor (IF) has been retrieved from the journal's page and its value reflects the average number of citations of articles recently published in the journal.

The journals and conventions vary from topics on artificial intelligence, bioinformatics, pattern recognition, biometrics, computer forensics, and security. All of this information is relevant and helps future researchers search for papers in the area or prospect in which journals and events have an affinity with the theme.

\section{Competitions in eye movement biometrics}

Although the competitions were not explicitly as a result of the search in the Scopus database, competitions boosted research in the area, with the aim to bring new researchers to the field and to collaborate with the results obtained over the years. Competitions have been identified through the titles of the articles collected.

The first competition was called EMVIC (Eye Movement Verification and Identification Competition)\footnote{http://www.emvic.org/} and was held in 2012 at the IEEE 5th International Conference on Biometrics: Theory Applications and Systems (BTAS) \cite{kasprowski2012first} in partnership with Kaggle website\footnote{http://www.kaggle.com/c/emvic}.  In this competition, 49 registered teams and 524 submissions were registered. The work of Cuong et al. \cite{cuong2012mel} extracted the eye features from eye raw data: cepstral coefficient, eye difference and eye velocity and, applied Decision tree, Bayesian network and Support vector machine and the results were an  Identification Rate (IR) of 93.56\% in Dataset A and 90.43\% in Dataset B. 



A second competition was launched, called EMVIC 2014 \cite{6996285}, and different from the first competition, used face images as a stimulus. The competition released 837 samples from 34 volunteers to the training dataset and 593 samples from 22 volunteers for the test dataset. There were 82  participants enrolled and 19 submissions, the best accuracy obtained was 39.63\%. The result was very different from the first competition with lower submissions and a smaller dataset, which could be affected by the results.

Another competition with the same objective was launched one year later, called BioEye 2015\footnote{https://bioeye.cs.txstate.edu/}  \cite{komogortsev2015bioeye}. The competition had 64 competitors, a total of 200 submissions along 26 days. In this edition were released four datasets to evaluate different parameters of the algorithms. 
 
The results obtained in BioEye 2015 demonstrated the potential of identification through eye movements, obtained in different scenarios \cite{rigas2017current}. Submissions were evaluated using the Rank-1 Identification Rate (IR), defined by the ratio of correct classifications with the amount of data. 



\section{Conclusion}

This paper presented a bibliometric analysis of eye movement biometrics based on the selected papers in journals and conferences between the years 2004 to 2019 indexed by Scopus. An interesting point was the competitions that helped to grow more and more research in the area. The number of publications increased significantly in the competition years (2012, 2014 and 2015) helping the authors to make partnerships and continuous studies. 
Using statistical methods of bibliometry the work has able to identify the network of collaboration between institutions and authors, most relevant articles, journals and conventions. 

\bibliographystyle{ijcaArticle}
\bibliography{Template}

\end{document}